\documentstyle [a4]{article}
\newcommand{\lab}{\label}

\newcommand{\bc}{\begin{center}}
\newcommand{\ec}{\end{center}}
\newcommand{\be}{\begin{equation}}
\newcommand{\ee}{\end{equation}}
\newcommand{\bea}{\begin{eqnarray}}
\newcommand{\eea}{\end{eqnarray}}
\newcommand{\bs}{\begin{subequations}}
\newcommand{\es}{\end{subequations}}
\newcommand{\beq}{\begin{eqalignno}}
\newcommand{\eeq}{\end{eqalignno}}

\def\lab{\label}

\def\lab{\label}

\begin{document}

\vspace{2.0cm}
\bc
\Large{\bf Quantum models related to fouled Hamiltonians of the harmonic
 oscillator}

\vspace{1.2cm}

\large{P. Tempesta, E. Alfinito$^{a)}$, R.A. Leo and G. Soliani} \\
\small
{\it Dipartimento di Fisica dell'Universit\`{a} di Lecce,\\
73100 Lecce, Italy, and Istituto Nazionale di Fisica Nucleare,\\
Lecce\\
 ${}^{a)}$ Dipartimento di Fisica dell'Universit\`{a} di Salerno,\\
84100 Baronissi, Salerno, Italy}

\vspace{1.5cm}

\ec

\vspace{1.2cm}

\small

{\bf Abstract}
We study a pair of canonoid (fouled) Hamiltonians of the harmonic oscillator
which provide, at the classical level, the same equation of motion as the
conventional Hamiltonian. These Hamiltonians, say $K_{1}$ and $K_{2},$
result to be explicitly time-dependent and can be expressed as a formal
rotation of two cubic polynomial functions, $H_{1}$ and $H_{2},$ of the
canonical variables $(q,p).$

We investigate the role of these fouled Hamiltonians at the quantum level.
Adopting a canonical quantization procedure, we construct some quantum
models and analyze the related eigenvalue equations. One of these models is
described by a Hamiltonian admitting infinite self-adjoint extensions, each
of them has a discrete spectrum on the real line. A self-adjoint extension
is fixed by choosing the spectral parameter $\varepsilon $ of the associated
eigenvalue equation equal to zero. The spectral problem is discussed in the
context of three different representations. For $\varepsilon =0,$ the
eigenvalue equation is exactly solved in all these representations, in which
square-integrable solutions are explicity found. A set of constants of
motion corresponding to these quantum models is also obtained. Furthermore,
the algebraic structure underlying the quantum models is explored. This
turns out to be a nonlinear (quadratic) algebra, which could be applied for
the determination of approximate solutions to the eigenvalue equations.

\normalsize

\section{ Introduction}

A few years ago, in Ref. [1] a method was devised to find alternative
Lagrangians for the time dependent oscillator
\begin{equation}
\stackrel{..}{q}+\omega (t)^{2}q=0,  \label{profsol}
\end{equation}
where $q=q(t),$ $\omega (t)$ is a given differentiable function, and the dot
stands for time-derivative. The method, which is based on the concept of
fouling transformation \cite{2}, is reviewed in Section II.

In this paper we study certain nonconventional quantum Hamiltonians
corresponding to the classical fouled Hamiltonians associated with the
Lagrangians derived in Ref. [1]. This investigation is motivated by the fact
that these nonconventional quantum Hamiltonians, having a polynomial
structure in the operators $a$ and $a^{\dagger },$ may play an important
role in the context of quantum optics especially in handling coherence and
squeezing of multiphoton systems.

All the fouled Lagrangians of the hierarchy found in Ref. [1] lead to the
same equation of motion (1) as it occurs for the conventional Lagrangian $%
L_{1}=\frac{1}{2}(\stackrel{.}{q}^{2}-\omega (t)^{2}q^{2})$ (see (5)). In
Section III the fouled Hamiltonians $K_{\pm},$ related to the
simplest fouled Lagrangians, $L_{2}^{(1)}$ and $L_{2}^{(2)},$ given by (23)
and (24) are written down. By way of example, we have limited ourselves to
consider the standard (harmonic) oscillator where $\omega (t)\equiv \lambda $
is a constant. Furthermore, it is shown that (at the classical level), as
one expects, $K_{\pm }$ reproduce the same equation of motion coming from
the conventional Hamiltonian $H_{0}=\frac{1}{2}(p^{2}+\lambda ^{2}q^{2}$ $)$%
. For brevity, we shall deal with $K_{+}$ only. It turns out that $K_{+}$
(see Section IV) can assume two possible forms, denoted by $K_{+}^{(1)}$ $%
\equiv K_{1}$ and $K_{+}^{(2)}$ $\equiv K_{2}$ (see (39) and (40)), which
are independent but depend explicitly on time via the coefficients $\cos
\lambda t$ and $\sin \lambda t.$ Consequently, $K_{1}$ and $K_{2}$ can be
formally interpreted as the result of a \textit{rotation} (of an angle $%
\lambda t$) of the quantities $H_{1}=\sqrt{\lambda }(p^{2}+\lambda
^{2}q^{2})q$ and $H_{2}=\frac{2}{3\sqrt{\lambda }}p^{3},$ which are \textit{%
not} explicitly dependent on time. In other words, we have $%
K_{1}^{2}+K_{2}^{2}=H_{1}^{2}+H_{2}^{2},$ and
\begin{equation}
\{K_{1},K_{2}\}_{q,p}=\{H_{1},H_{2}\}_{q,p}
\end{equation}
(see Section IV), where the symbol $\{,\}$ denotes the Poisson bracket with
respect to the canonical variables $(q,p),$ namely
\begin{equation}
\{A,B\}_{q,p}=\frac{\partial A}{\partial q}\frac{\partial B}{\partial p}-%
\frac{\partial A}{\partial p}\frac{\partial B}{\partial q}.
\end{equation}
In Section V the canonical quantization procedure based on a bosonic couple
of annihilation and creation operators, $a$ and $a^{\dagger }$ (with $%
[a,a^{\dagger }]=1),$ is applied to the fouled Hamiltonians $K_{1}$ and $%
K_{2}.$ In such a way $K_{1}$ and $K_{2}$ are converted into a pair of
Hamiltonian operators, say ${\cal K}_{1}$ and ${\cal K}_{2}.$
(Classically, as we know, $K_{1}$ and $K_{2}$ provide the same equation of
motion (1)). Furthermore, since the canonical quantization procedure does
affect only the cubic polynomial functions of $(q,p),$ namely $H_{1}$ and $%
H_{2},$ here we limit ourselves to study the operators ${\cal H}_{1}$ and 
${\cal H}_{2},$ which are the quantized versions of $H_{1}$ and $H_{2},$
respectively. The arising \textit{quantum models} are investigated in
Sections VI and VII.

Precisely, in Section VI we show that the quantum model coming from $%
{\cal H}_{1}$ gives rise to a linear second order eigenvalue equation of
the Sturm-Liouville type (in the harmonic oscillator excitation number
representation, or $n-$rep).

An interesting feature of the operator ${\cal H}_{1}$ is that it is
connected with an undetermined Hamburger moment problem \cite{3,4,5}. We
show that this operator has deficiency indices (1,1) and allows a
one-parameter family of self-adjoint extensions whose spectra are discrete
and with no point in common. This goal has been achieved essentially by
associating the operator ${\cal H}_{1}$ with a Jacobi matrix \cite{3,5}.
A possible physical interpretation of the operator ${\cal H}_{1}$ is
provided by fixing, among the infinite self-adjoint extensions, the
extension corresponding to $\varepsilon =0,$ where $\varepsilon $ denotes
the spectral parameter. In this case the eigenvalue equations for ${\cal H%
}_{1}$ can be solved exactly in all the following representations: the
harmonic oscillator excitation number representation ($n-$ rep), the
coordinate space representation ($q-$ rep), and the Fock-Bargmann
holomorphic function representation ($z-$ rep) \cite{6,7,8}. In all these
representations, for $\varepsilon =0$ square-integrable solutions of the
eigenvalue equations are explicitly obtained.

In Section VII we deal with the eigenvalue equation for ${\cal H}_{2}.$
The analysis of this Hamiltonian is trivial. It is exactly solvable \cite{9}%
, its spectrum is on the whole line and the related eigenfunctions can be
easily found. Nevertheless, ${\cal H}_{2}$ has been involved in the
construction of two quantum models described by the operators ${\cal H}%
_{3}$ and ${\cal H}_{4}$ (see (57) and (58)), where ${\cal H}_{2}$ $=$ 
${\cal H}_{3}$ $+$ ${\cal H}_{4}.$ One can attribute to ${\cal H}%
_{3}$ a physical meaning. Specifically, ${\cal H}_{3}$ can be interpreted
as a special case of a class of Hamiltonians appearing in the higher order
nonlinear optical processes \cite{5,10,11,12}. This Hamiltonian is involved
in the construction of third power squeezed states \cite{5,11,13}.

On the other hand, as we can see in Section VII, the operator ${\cal H}%
_{4}$ is closely related to ${\cal H}_{1}$ so that the solutions of the
corresponding eigenvalue equation can be derived from the solutions of the
eigenvalue equation for ${\cal H}_{1}$ .

In Section VIII some constants of motion involving the operators ${\cal H}%
_{1}$,${\cal H}_{2},{\cal H}_{3},{\cal H}_{4}$ and ${\cal H}_{5}$
$\sim a^{3}+a^{\dagger 3}$ are derived. Section IX contains a discussion on
a possible algebraic framework which could be employed to analyze Eq. (69).
This approach resorts to a quadratic algebra in terms of which the operator $%
{\cal H}_{1}$ can be naturally expressed. Finally, in Section X a few
concluding remarks are presented, while in the Appendix a detail of a
calculation is reported.

\bigskip

\section{\protect\bigskip Fouled Lagrangians}

\bigskip

We recall that a fouling transformation is a transformation under which the
coordinates in configuration space are preserved:
\begin{equation}
Q=q,
\end{equation}
while
\begin{equation}
P_{n}=P_{n}(q,p,t),
\end{equation}
$P_{n}$ being a polynomial of degree $n$ in the variables $q$ and $p=\frac{%
\partial L_{1}}{\partial \stackrel{.}{q}}=\stackrel{.}{q},$ i.e.
\begin{equation}
P_{n}=\sum^{n}_{0}\, a_{j}p^{n-j}q^{j},
\end{equation}
where
\begin{equation}
L_{1}=\frac{1}{2}(\stackrel{.}{q^{2}}-\omega (t)^{2}q^{2})
\end{equation}
is the conventional Lagrangian, and $a_{j}=a_{j}(t)$ are (real)
time-dependent coefficients.

In \cite{1} it was proven that the function $L_{n}$ $=L_{n}(q,\stackrel{.}{q}%
,t)$ expressed by
\begin{equation}
L_{n}=
\sum^{n}_{j=0}\frac{1}{n-j+1}a_{j}\stackrel{.}{q}%
^{n-j+1}q^{j}+(\stackrel{.}{a}_{n}-\omega ^{2}a_{n-1})\frac{q^{n+1}}{n+1}
\end{equation}
satisfies the equations
\begin{equation}
\frac{\partial L_{n}}{\partial \stackrel{.}{q}}=P_{n},\quad\frac{\partial
L_{n}}{\partial q}=\stackrel{.}{P}_{n}.
\end{equation}
The compatibility condition for these equations provides
\begin{equation}
\stackrel{.}{a}_{o}=-\frac{n-1}{n}a_{1},\quad\stackrel{.}{a}%
_{j}=(n-j+1)\omega ^{2}a_{j-1}-(j+1)\frac{n-j-1}{n-j}a_{j+1},
\end{equation}
with $j=1,2,...,n-1.$ Furthermore, we have
\begin{equation}
\frac{d}{dt}\frac{\partial L_{n}}{\partial \stackrel{.}{q}}-\frac{\partial
L_{n}}{\partial q}=I_{n}(t)(\stackrel{..}{q}+\omega (t)^{2}q),
\end{equation}
where $I_{n}(t)$ is a time-dependent constant of motion, viz. $\frac{d}{dt}%
I_{n}(t)$ $=0,$ given by
\begin{equation}
I_{n}(t)=
\sum^{n}_{j=0}(n-j)a_{j}p^{n-j-1}q^{j}=\frac{%
\partial ^{2}L_{n}}{\partial \stackrel{.}{q}^{2}},
\end{equation}
for any $n\geq 1.$

From Eq. (9) we deduce that corresponding to the solution $q$ of the
generalized oscillator (1), $L_{n}$ satisfies the Euler-Lagrange equation.
We notice that for $n=1$, Eq. (6) gives the conventional Lagrangian (5)
(with $a_{0}=1$ and $a_{1}=0).$ The related invariant is $I_{1}=\frac{%
\partial ^{2}L_{1}}{\partial \stackrel{.}{q}^{2}}=1.$

Hereafter, by way of example, we are interested in the case $n=2.$ Then,
from (6) and (10) we obtain
\begin{equation}
L_{2}=\frac{1}{3}a_{0}\stackrel{.}{q}^{3}+\frac{1}{2}a_{1}\stackrel{.}{q}%
^{2}q+a_{2}\stackrel{.}{q}q^{2}+\frac{1}{3}(\stackrel{.}{a_{2}}-\omega
^{2}a_{1})q^{3},
\end{equation}
\begin{equation}
I_{2}=2a_{0}p+a_{1}q,
\end{equation}
where (see (8))
\begin{equation}
\stackrel{.}{a}_{0}=-\frac{1}{2}a_{1},\quad\stackrel{.}{a}_{1}=2\omega
^{2}(t)a_{0},
\end{equation}
namely
\begin{equation}
\stackrel{..}{a}_{0}+\omega ^{2}(t)a_{0}=0.
\end{equation}
The general solution of Eq. (14) can be written in the form
\begin{equation}
a_{0}=\sqrt{2}\sigma (c_{1}\cos \frac{\theta }{2}+c_{2}\sin \frac{\theta }{2}%
),
\end{equation}
where $c_{1},c_{2}$ are constants, and $\sigma ,\theta $ are defined by
\begin{equation}
\stackrel{..}{\sigma }+\omega ^{2}(t)\sigma =\frac{1}{4\sigma ^{3}},
\end{equation}
and
\begin{equation}
\stackrel{.}{\theta }=\frac{1}{\sigma ^{2}}.
\end{equation}
In the following we shall limit ourselves to the choice $\omega (t)=\lambda
=const.$ Consequently, Eqs. (16) and (17) admit the solution
\begin{equation}
\sigma =\frac{1}{\sqrt{2\lambda }}
\end{equation}
and
\begin{equation}
\theta =2\lambda t.
\end{equation}
Equation (15) has the two independent solutions
\begin{equation}
a_{0}^{(1)}=\frac{1}{\sqrt{\lambda }}\cos \lambda t,\quad a_{0}^{(2)}=%
\frac{1}{\sqrt{\lambda }}\sin \lambda t,
\end{equation}
while the corresponding values $a_{1}^{(1)},$ $a_{1}^{(2)}$ are (see (13))
\begin{equation}
a_{1}^{(1)}=2\sqrt{\lambda }\sin \lambda t,\quad a_{1}^{(2)}=-2\sqrt{%
\lambda }\cos \lambda t.
\end{equation}
On the other hand, the expressions of $a_{2}^{(1)},$ $a_{2}^{(2)}$ turn out
to be (see \cite{1})
\begin{equation}
a_{2}^{(1)}=\lambda ^{\frac{3}{2}}\cos \lambda t,\quad a_{2}^{(2)}=\lambda
^{\frac{3}{2}}\sin \lambda t.
\end{equation}
Now, using (20), (21) and (22), from (11) we obtain the two alternative
fouled Lagrangians
\begin{equation}
L_{2}^{(1)}=\frac{1}{3\sqrt{\lambda }}\cos \lambda t\stackrel{.}{q}^{3}+\sqrt{%
\lambda }\stackrel{}{\sin \lambda t\quad \stackrel{.}{q^{2}}}q+\lambda ^{^{%
\frac{3}{2}}}\cos \lambda t\stackrel{.}{q}q^{2}-\lambda ^{\frac{5}{2}}\sin
\lambda tq^{3},
\end{equation}
\begin{equation}
L_{2}^{(2)}=\frac{1}{3\sqrt{\lambda }}\sin \lambda t\stackrel{.}{q}^{3}-\sqrt{%
\lambda }\stackrel{}{\cos \lambda t\quad \stackrel{.}{q^{2}}}q+\lambda ^{%
\frac{3}{2}}\sin \lambda t\stackrel{.}{q}q^{2}+\lambda ^{\frac{5}{2}}\cos
\lambda tq^{3},
\end{equation}
which furnish the two independent invariants
%
\begin{equation}
I_{2}^{(1)}=\frac{\partial ^{2}L_{2}^{(1)}}{\partial \stackrel{.}{q}^{2}}=%
\frac{2}{\sqrt{\lambda }}\cos \lambda t\stackrel{.}{q}+2\sqrt{\lambda }\sin
\lambda t\quad q,
\end{equation}
\begin{equation}
I_{2}^{(2)}=\frac{\partial ^{2}L_{2}^{(2)}}{\partial \stackrel{.}{q}^{2}}=%
\frac{2}{\sqrt{\lambda }}\sin \lambda t\stackrel{.}{q}-2\sqrt{\lambda }\cos
\lambda t\quad q,
\end{equation}
respectively (see (10)).

Equations (25) and (26) provide the general solution $q$ and the momentum $p=%
\stackrel{.}{q}$ of the harmonic oscillator ($\omega \equiv \lambda ),$ i.e.

\begin{equation}
q=\frac{1}{2\sqrt{\lambda }}(I_{2}^{(1)}\sin \lambda t-I_{2}^{(2)}\cos
\lambda t),
\end{equation}
\begin{equation}
p=\stackrel{.}{q}=\frac{\sqrt{\lambda }}{2}(I_{2}^{(1)}\cos \lambda
t+I_{2}^{(2)}\sin \lambda t),
\end{equation}
from which, as one expects,
\begin{equation}
H_{0}=\frac{1}{2}(p^{2}+\lambda ^{2}q^{2})=\frac{\lambda }{8}[%
(I_{2}^{(1)})^{2}+(I_{2}^{(2)})^{2}]=const,
\end{equation}
where $H_{0}$ is the conventional Hamiltonian.


\section{Fouled Hamiltonians}


At this point, we build up the Hamiltonian corresponding to the Lagrangian $%
L_{2}$ given by (11). Following the procedure of Ref. \cite{1}, we get
\begin{equation}
K_{\pm }=-\frac{a_{1}}{2a_{0}}qP\pm \frac{2}{3}a_{0}[(\frac{a_{1}^{2}}{%
4a_{0}^{2}}-\frac{a_{2}}{a_{0}})q^{2}+\frac{P}{a_{0}}]^{\frac{3}{2}}+[\frac{%
a_{1}}{6a_{0}}(3a_{2}-\frac{a_{1}^{2}}{2a_{0}})-\frac{1}{3}(\stackrel{.}{a_{2}%
}-\lambda ^{2}a_{1})]q^{3},
\end{equation}
where
\begin{equation}
P=a_{0}p^{2}+a_{1}qp+a_{2}q^{2}.
\end{equation}
For the sake of definiteness, later we shall study the Hamiltonian $K_{+}$
only.

The Hamilton equations for $K_{+}$ read
\begin{equation}
\stackrel{.}{q}=-\frac{a_{1}}{2a_{0}}q+\frac{I_{2}}{2a_{0}},
\end{equation}
\begin{equation}
\stackrel{.}{P}=\frac{a_{1}}{2a_{0}}P-\frac{a_{1}^{2}-4\lambda ^{2}a_{0}^{2}}{%
4a_{0}^{2}}I_{2}q-3a_{1}(\lambda ^{2}-\frac{a_{1}^{2}}{12a_{0}^{2}})q^{2},
\end{equation}
where the relation
\begin{equation}
(\frac{a_{1}^{2}}{4a_{0}^{2}}-\frac{a_{2}}{a_{0}})q^{2}+\frac{P}{a_{0}}=(%
\frac{I_{2}}{2a_{0}})^{2}
\end{equation}
with $a_{2}=\lambda ^{2}a_{0}$ has been used (see \cite{1}).

By expliciting Eqs. (31) and (32) with the help of (12) and (30), we arrive
at the equation
\begin{equation}
I_{2}(\stackrel{..}{q}+\lambda ^{2}q)=0,
\end{equation}
which tells us that the fouled Hamiltonian $K_{+},$ as it occurs for the
related fouled Lagrangian $L_{2}$ (see (9)), gives rise to the same equation
of motion emerging from the conventional Hamiltonian (29).

By virtue of (33), the Hamiltonians (30) take the form
\begin{equation}
K_{+}=\frac{1}{2}a_{1}(p^{2}+\lambda ^{2}q^{2})q+\frac{2}{3}a_{0}p^{3},
\end{equation}
\begin{equation}
K_{-}=-\frac{3}{2}a_{1}p^{2}q-\frac{a_{1}^{2}}{a_{0}}q^{2}p+(\frac{a_{1}}{2}%
\lambda ^{2}-\frac{a_{1}^{3}}{6a_{0}^{2}})q^{3}-\frac{2}{3}a_{0}p^{3}.
\end{equation}
One can see that
\begin{equation}
K_{\pm }=Pp-L_{2},
\end{equation}
where $P$ and $L_{2}$ are given by (31) and (11).

\section{The fouled Hamiltonians $K_{+}^{(1)}$ and $K_{-}^{(2)}$}

By using (20), (21) and (22), from (36) we derive the pair of (explicitly
time-dependent) Hamiltonians
\begin{equation}
K_{+}^{(1)}\equiv K_{1}=H_{1}\sin \lambda t+H_{2}\cos \lambda t,
\end{equation}
\begin{equation}
K_{+}^{(2)}\equiv K_{2}=-H_{1}\cos \lambda t+H_{2}\sin \lambda t,
\end{equation}
where $H_{1}$ and $H_{2}$ are defined by
\begin{equation}
H_{1}=\sqrt{\lambda }(p^{2}+\lambda ^{2}q^{2})q=2\sqrt{\lambda }H_{0}q,
\end{equation}
\begin{equation}
H_{2}=\frac{2}{3\sqrt{\lambda }}p^{3}.
\end{equation}
One can check straightforwardly that the following evolution equations
\begin{equation}
\stackrel{.}{K_{1}}=\{K_{1},H_{0}\}_{q,p}+\frac{\partial K_{1}}{\partial t},
\end{equation}
\begin{equation}
\stackrel{.}{K_{2}}=\{K_{2},H_{0}\}_{q,p}+\frac{\partial K_{2}}{\partial t},
\end{equation}
\begin{equation}
\stackrel{.}{H_{1}}=\{H_{1},H_{0}\}_{q,p},
\end{equation}
\begin{equation}
\stackrel{.}{H_{2}}=\{H_{2},H_{0}\}_{q,p},
\end{equation}
hold, where $H_{0}$ is given by (29) and
\begin{equation}
\{H_{1},H_{0}\}_{q,p}=2\sqrt{\lambda }pH_{0},
\end{equation}
\begin{equation}
\{H_{2},H_{0}\}_{q,p}=-2\frac{\lambda ^{2}}{\sqrt{\lambda }}p^{2}q.
\end{equation}
We also have that the Poisson brackets between $K_{1},K_{2}$ and $%
H_{1},H_{2} $ concide, i.e.
\begin{equation}
\{K_{1},K_{2}\}_{q,p}=\{H_{1},H_{2}\}_{q,p}.
\end{equation}
This is a direct consequence of the rotation form of the transformations
(39) and (40).

\section{Quantization}

Hereafter, we shall put for simplicity $\lambda =\hbar =1.$

In order to quantize the fouled Hamiltonians (39) and (40), let us introduce
the operators
\begin{equation}
\stackrel{\wedge }{q}=\frac{1}{\sqrt{2}}(a+a^{\dagger }),\quad \stackrel{%
\wedge }{p}=-\frac{i}{\sqrt{2}}(a-a^{\dagger }),
\end{equation}
where $a$ and $a^{\dagger }$ denote a (boson) annihilation and a creation
operator, respectively.

By means of (50), we can write the operators ${\cal H}_{1}$and ${\cal H%
}_{2}$ corresponding to the classical functions (41) and (42). We obtain
\begin{equation}
{\cal H}_{1}=\sqrt{2}(a^{\dagger 2}a+a^{\dagger }a^{2}+a^{\dagger }+a),
\end{equation}
\begin{equation}
{\cal H}_{2}=\frac{i}{\sqrt{2}}[\frac{1}{3}(a^{3}-a^{\dagger
3})+(a^{\dagger 2}a-a^{\dagger }a^{2}+a^{\dagger }-a)],
\end{equation}
with the help of the commutation rule
\begin{equation}
\lbrack a,a^{\dagger }]=1
\end{equation}
or, in terms of the Heisenberg commutation relation: $[\stackrel{\wedge }{q},%
\stackrel{\wedge }{p}]=i.$

\qquad Now let us use the representation $\stackrel{\wedge }{q}=x,$ $\stackrel{%
\wedge }{p}=-i\frac{d}{dx},$ so that the operators $a$ and $a^{\dagger }$
can be written as (see (50))
\begin{equation}
a=\frac{1}{\sqrt{2}}(x+\frac{d}{dx}),\quad a^{\dagger }=\frac{1}{\sqrt{2}}%
(x-\frac{d}{dx}).
\end{equation}
In terms of these quantities, ${\cal H}_{1}$and ${\cal H}_{2}$ take
the forms
%
\begin{equation}
{\cal H}_{1}\quad =-(x\frac{d^{2}}{dx^{2}}+\frac{d}{dx})+x^{3},
\end{equation}
%

and
\begin{equation}
{\cal H}_{2}=\frac{2i}{3}\frac{d^{3}}{dx^{3}}.
\end{equation}
For later convenience, we shall report also in the representation (54) the
following operators appearing in (52):

\begin{equation}
{\cal H}_{3}\equiv \frac{i}{\sqrt{2}}\frac{1}{3}(a^{3}-a^{\dagger 3})=%
\frac{i}{2}[\frac{1}{3}\frac{d^{3}}{dx^{3}}+(x^{2}\frac{d}{dx}+x)],
\end{equation}
\begin{equation}
{\cal H}_{4}\equiv \frac{i}{\sqrt{2}}(a^{\dagger 2}a-a^{\dagger
}a^{2}+a^{\dagger }-a)=\frac{i}{2}[\frac{d^{3}}{dx^{3}}-(x^{2}\frac{d}{dx}+x)%
].
\end{equation}
In the next Section, we shall study the operators ${\cal H}_{1}$and $%
{\cal H}_{2}$ by dealing with the corresponding eigenvalue problems.

\section{ Self-adjoint extensions of the operator 
${\cal H}_{1}$}


In order to clarify the quantum-mechanical meaning of the Hamiltonian $%
{\cal H}_{1},$ it is crucial to establish whether ${\cal H}_{1}$
enjoys self-adjoint type properties. In doing so, first let us recall that
one can provide different representations for the operator (51), which
correspond to various forms of the related eigenvalue equations. We shall
consider the following representations: the harmonic oscillator excitation
number representation ($n-$ rep), the coordinate representation ($q-$ rep),
and the Fock-Bargmann holomorphic function reprentation ($z-$ rep).

For reader's convenience, we shall summarize below the main properties of
these representations \cite{8}.

Let us $\bf{H}$ denote the Hilbert space where the operators $\stackrel{%
\wedge }{q},$ $\stackrel{\wedge }{p}$ and $a^{\dagger },$ $a$ act. In the $n-$
rep, the vectors $\{\mid n>\}$ form a basis in $\bf{H}$. The following
relations
\begin{equation}
a=\sqrt{n}\mid n-1>,\quad a^{\dagger }\mid n>=\sqrt{n+1}\mid n+1>,\quad %
a^{\dagger }a\mid n>=n\mid n>
\end{equation}
hold.

On the other hand, in the $q-$ rep a vector $\mid \psi >$ belonging to the
space $\bf{H}$ is represented by a coordinate function $<q\mid \psi >=\psi
(q)$ which is square-integrable:
\begin{equation}
\int_{-\infty }^{+\infty }\mid \psi (q)\mid ^{2}dq<\infty .
\end{equation}
The basic vector $\mid n>$ is described by the function
\begin{equation}
<q\mid n>=\varphi _{n}(q)=N_{n}H_{n}(q)\exp (-\frac{q^{2}}{2}),
\end{equation}
where $N_{n}=(\sqrt{\pi }2^{n}n!)^{-\frac{1}{2}}$ and $H_{n}(q)$ is the
Hermite polynomial of degree $n.$ In the $q-$ rep, formulas (59) become the
standard recursion relations for the Hermite polynomials.

In order to introduce the Fock-Bargmann representation (or $z-$ rep), let $%
\mid \psi >$ be an arbitrary normalized vector in $\bf{H},$ namely
\begin{equation}
\mid \psi >=
\sum^{\infty}_{n=0}c_{n}\mid n>,
\end{equation}
with $<\psi \mid \psi >=
\sum^{\infty}_{n=0}\mid
c_{n}\mid ^{2}=1.$ Furthermore, taking into account the Glauber form
\begin{equation}
\mid z>=\exp (-\frac{\mid z\mid ^{2}}{2})
\sum^{\infty}_{n=0}\frac{z^{n}}{\sqrt{n!}}\mid n>,
\end{equation}
the state $\mid \psi >$ is completely determined by
\begin{equation}
<z\mid \psi >=\exp (-\frac{\mid z\mid ^{2}}{2})\psi (\stackrel{-}{z}),
\end{equation}
where
\begin{equation}
\psi (z)=\sum^{\infty}_{n=0}c_{n}u_{n}(z),
\end{equation}
with $u_{n}(z)=\frac{z^{n}}{\sqrt{n!}}.$ Owing to the condition 
$\sum^{\infty}_{n=0}\mid c_{n}\mid ^{2}=1,$ the series in (65)
converges uniformly in any compact domain of the complex $z$ plane.
Consequently, $\psi (z)$ turns out to be an entire holomorphic function in
the $z$ plane, and
\begin{equation}
\parallel \psi \parallel ^{2}=\int^{{}}\exp (-\mid z\mid ^{2})\mid
\psi (z)\mid ^{2}d\mu (z)<\infty ,
\end{equation}
where $d\mu (z)=\pi ^{-1}dxdy,$ $z=x+iy$ \cite{8}.

The scalar product of two entire functions $\psi _{1}(z)$ and $\psi _{2}(z)$
obeying the condition (66) is given by
\begin{equation}
<\psi _{1}\mid \psi _{2}>=\int \exp (-\mid z\mid ^{2})\stackrel{-}{\psi }%
_{1}(z)\psi _{2}(z)d\mu (z).
\end{equation}
As it was proven by Bargmann \cite{7}, the Fock-Bargmann representation
space with a scalar product provided by (67), is really a Hilbert space. We
observe also that in the $z-$rep, the operator solution for the commutation
relation $[a,a^{\dagger }]=1$ \cite{6} is
\begin{equation}
a\rightarrow \frac{d}{dz},\quad a^{\dagger }\rightarrow z.
\end{equation}
Now let us consider the eigenvalue equation
\begin{equation}
(a^{\dagger 2}a+a^{\dagger }a^{2}+a+a^{\dagger })\mid \chi >\quad %
=\varepsilon \mid \chi >.
\end{equation}
%
Starting from the $n-$ rep and following the lines of Ref. \cite{5}, 
let us put
\begin{equation}
\mid \chi >=\sum_{n=0}f_{n}(\varepsilon )\mid n>
\end{equation}
into Eq. (69). By using Eqs. (59), after simple calculations we obtain the
recursion formula
\begin{equation}
(n+1)^{\frac{3}{2}}f_{n+1}(\varepsilon )-\varepsilon f_{n}(\varepsilon )+n^{%
\frac{3}{2}}f_{n-1}(\varepsilon )=0
\end{equation}
for $n\geq 1,$ and
\begin{equation}
f_{1}(\varepsilon )=\varepsilon f_{0}(\varepsilon )
\end{equation}
(the boundary condition).

We remark that the sequence $\{f_{n}\}$ is such that the series $\sum_{n=0}
\mid f_{n}\mid ^{2}$ converges, i.e. the sequence $%
\{f_{n}\}\equiv (f_{0},f_{1},f_{2},...)$ belongs to the Hilbert space $%
l^{2}. $

To show this, first let us consider the case $\varepsilon =0.$ Then, Eqs.
(71)-(72) provide
\begin{equation}
f_{2n}=(-1)^{n}[\frac{(2n-1)!!}{(2n)!!}]^{\frac{3}{2}}f_{0},
\end{equation}
the odd terms being zero. Thus, from the asymptotic formula of Gamma
function ([14], p. 257). we deduce that for $n\rightarrow \infty ,$ $\mid
f_{2n}\mid ^{2}$ behaves as $n^{-\frac{3}{2}},$ so that $\sum_{n=0}\mid f_{2n}\mid ^{2}<\infty .$ In general, i.e. for $\varepsilon \neq 0,$
Eq. (71) tells us that for great values of $n,$ the sum of the first and the
last term is leading with respect to the second term. This allows us to see
easily that both even and odd terms, $\mid f_{2n}\mid ^{2}$ and $\mid
f_{2n+1}\mid ^{2},$ behave asymptotically as $n^{-\frac{3}{2}}.$ To
conclude, the series $\sum_{n=0}\mid f_{n}\mid ^{2}$ is
convergent, namely the sequence $\{f_{n}\}$ belongs to the Hilbert space 
$l^{2}$ for any value of the spectral parameter $\varepsilon .$ Then, the
equation ${\cal H}_{1}^{\dagger }\mid \chi >=\varepsilon \mid \chi >,$
for $\Im \lambda \neq 0,$ has nontrivial solutions ([3], p. 140).

Now, by introducing the notation $b_{n}=(n+1)^{\frac{3}{2}},$ we see that it
is possible to associate with the difference equation (71) the (infinite)
Jacobi matrix

\begin{equation}
A=\left( 
\begin{array}{ccccc}
0 & b_{0} & 0 & 0 & ... \\ 
b_{0} & 0 & b_{1} & 0 & ... \\ 
0 & b_{1} & 0 & b_{2} & ... \\ 
0 & 0 & b_{2} & 0 & ... \\ 
.. & ... & ... & ... & ...
\end{array}
\right)
\end{equation}
so that Eq. (69) is equivalent to the eigenvalue equation $Af=\varepsilon f,$
with $f=(f_{0},f_{1},$ $...)^{T}.$

The Jacobi matrix (74) plays a crucial role in the study of the Hamburger
moment problem (see, for example, \cite{3,5,15}). Precisely, let us consider
the moments
\begin{equation}
s_{n}=\int_{-\infty }^{+\infty }x^{n}d\sigma (x),\quad n=0,1,2,...,
\end{equation}
where $\sigma $ denotes a (positive) measure on $\cal R$ (\cite{15}, p.
145). The Hamburger moment problem is to determine conditions on a sequence
of real numbers $\stackrel{}{\{s_{n}\}_{n=0}^{\infty }}$ , so that there
exists a measure satisfying (75). One can show that a sequence of real
numbers $\{s_{n}\}$ are the moments of a positive measure on $\cal R$ if
and only if for all $N$ and all $\alpha _{0},\alpha _{1},...,\alpha _{N}$ $%
\in \cal{C}$, one has
\begin{equation}
\sum^{N}_{n,m=0}\stackrel{-}{\alpha }_{n}\alpha
_{m}s_{n+m}\geq 0.
\end{equation}
%
From the Jacobi matrix (74) we get the limitations
\begin{equation}
\sum^{\infty}_{n=0}\frac{1}{b_{n}}=
\sum^{\infty}_{n=0}\frac{1}{(n+1)^{\frac{3}{2}}}<\infty ,\quad %
b_{n-1}b_{n+1}<b_{n}^{2}.
\end{equation}
Consequently, the Jacobi matrix (74) belongs to the $type$ $C$ (limit circle
case), and corresponds to an undetermined Hamburger moment problem (\cite
{3,5}). So, the properties of the operator ${\cal H}_{1}$ $\sim $ $%
a^{\dagger 2}a+a^{\dagger }a^{2}+a+a^{\dagger }$ are similar to the
properties of the operator $a^{k}+a^{\dagger k}$ ($k=3)$ discussed by Nagel 
\cite{5}. In other words, the operator ${\cal H}_{1}$ has deficiency
indices (1,1) and allows a one-parameter family of self-adjoint extensions,
each having a purely discrete spectrum on the real line \cite{4,16}. The
spectra of two different extensions turn out to have no point in common ( 
\cite{3}, p. 152). We have that different self-adjoint extensions correspond
to different dynamics \cite{15,17}.

Since every self-adjoint extension of ${\cal H}_{1}$ has a discrete
spectrum on the real line, and taking one eigenvalue determines the
corresponding extension uniquely, let us choose $\varepsilon =0.$

In this case the eigenvalue equation underlying the operator ${\cal H}%
_{1} $ can be solved exactly in all the representations mentioned at the
beginning of this Section.

To show this, let us deal with the $q-$ rep. With the help of (54), the
eigenvalue equation (69) reads
\begin{equation}
x\phi ^{\prime \prime }+\phi ^{\prime }+(\sqrt{2}\varepsilon -x^{3})\phi =0,
\end{equation}
where $\phi ^{\prime }\equiv \frac{d}{dx}\phi $.

Equation (78) can be written as the Sturm-Liouville equation (\cite{18}, p.
59)
\begin{equation}
{\cal L}[\phi (x)]{\cal =-}\sqrt{2}{\cal \varepsilon \phi (}x%
{\cal )},
\end{equation}
where ${\cal L}$ denotes the Sturm-Liouville operator
\begin{equation}
{\cal L=}\frac{d}{dx}[x\frac{d}{dx}]-x^{3}.
\end{equation}
By using the transformation
\begin{equation}
\phi (x)=\exp (-\frac{x^{2}}{2})\psi (x),
\end{equation}
Equation (78) becomes
\begin{equation}
x\psi ^{\prime \prime }+(1-2x^{2})\psi ^{\prime }+(\sqrt{2}\varepsilon
-2x)\psi =0.
\end{equation}
This equation is satisfied by
\begin{equation}
\psi (x)=\sum^{\infty}_{n=0}\ f_{n}\ N_{n}\ H_{n}(x),
\end{equation}
where the coefficients $f_{n}$ fulfil the recursion relations (71) and (72).
We have already shown that $\{f_{n}\}$ $\in l^{2}.$ Then, the function $\psi
(x)$ belongs to the Hilbert space $L_{e^{-x^{2}}}^{2}(-\infty ,\infty )$ ($%
\exp (-x^{2})$ is the weight function). We point out that generally the
relation (83) is not valid in the pointwise sense, but it holds in
accordance with the metric of $L_{e^{-x^{2}}}^{2}(-\infty ,\infty ),$ namely
\begin{equation}
\lim_{n\rightarrow \ \infty } \int_{-\infty }^{+\infty }\mid 
{\psi (x)-\sum^{n}_{k=0}} \ f_{k}\ N_{k}\ H_{k}(x)
\mid ^{2}\exp (-x^{2})dx=0.
\end{equation}
For $\varepsilon =0,$ via the change of variable $\xi =x^{2}$ Eq. (82) can
be written as a special case of a Kummer equation, whose independent
solutions are $M(\frac{1}{2},1;x^{2})$ and $\quad U(\frac{1}{2},1;x^{2})$ ( 
\cite{14}, p. 504). In the case the solution $\psi (x)$ of Eq. (82) with the
property $\psi (x)\in $ $L_{e^{-x^{2}}}^{2}(-\infty ,\infty )$ is given by
\begin{equation}
\psi (x)=c\ U(\frac{1}{2},1;x^{2})=
\sum^{\infty}_{n=0}\ f_{2n}\ N_{2n}\ H_{2n}(x),
\end{equation}
where the constant $c$ is such that $c\pi ^{\frac{3}{4}}=f_{0},$ and
\begin{equation}
f_{2n}=f_{0}\pi ^{-\frac{3}{4}}N_{2n}\int_{-\infty }^{+\infty }U(\frac{1}{2}%
,1;x^{2})H_{2n}(x)\exp (-x^{2})dx.
\end{equation}
One can easily check that (86) is satisfied for any $n\in \cal{N}$.

Finally, in the $z-$ rep, i.e. for $\mid \chi >$ $\rightarrow \chi ,$ $%
a^{\dagger }\rightarrow z,$ $a\rightarrow \frac{d}{dz},$ Eq. (69) gives
\begin{equation}
z\chi _{zz}+(1+z^{2})\chi _{z}+(z-\varepsilon )\chi =0,
\end{equation}
where
\begin{equation}
\chi =\sum^{\infty}_{n=0}f_{n}\frac{z^{n}}{\sqrt{n!}}
\end{equation}
and $f_{n}$ satisfies the recursion relations (71) and (72). For $%
\varepsilon =0,$ the eigenvalue equation (87) as well can be exactly solved.
In fact, by setting $z^{2}=y,$ $\zeta =-\frac{y}{2},$ this equation becomes
a special case of the Kummer equation.

Therefore, in the $z-$ rep, where the eigenfunction should be a holomorphic
(and normalizable) function in the whole $z-$ plane, one has the solution $M(%
\frac{1}{2},1;-\frac{z^{2}}{2}).$ The other solution, namely the Kummer
function $U(\frac{1}{2},1;-\frac{z^{2}}{2})$, is not holomorphic at $z=0.$

To conclude this Section, we observe that the solutions of the eigenvalue
equation for ${\cal H}_{1}$ in the case $\varepsilon =0$ can be also
found from Eqs. (71) and (72), obtained within the $n-$ rep, by means of the
standard integral representations of the confluent hypergeometric functions.
An example of this procedure is displayed in \cite{19}.

\section{The ${\cal H}_{2}$, ${\cal H}_{3}$, ${\cal H}_{4}$ models}


The eigenvalue problem for the operator ${\cal H}_{2}$ can be written as
\begin{equation}
{\cal L}\psi =\varepsilon \psi ,
\end{equation}
where ${\cal L=}\frac{2i}{3}D_{x}^{3}$ $(D_{x}=\frac{d}{dx}).$ The study
of Eq. (89) is trivial. It is exactly solvable \cite{9}, its spectrum is on
the whole line, and the related eigenfunctions are of the exponential type.

The operator ${\cal H}_{3}$ (see (57)) belongs to the class of
Hamiltonians
\begin{equation}
{\cal H=}\quad i\kappa _{n}(a^{n}-a^{\dagger n})
\end{equation}

appearing in the higher order nonlinear optical processes. In particular,
(57) describes a subharmonic generation process, in which a photon from a
strong pump beam produces $n$ photons of the signal beam in a nonlinear
medium \cite{11}. The constant $\kappa _{n}$ is related to the $n$th
nonlinear susceptibility coefficient and to the amplitude of the pump field,
while $a$ and $a^{\dagger }$ are the annihilation and the creation operators
for the signal field. In this context, the evolution of an arbitrary initial
state $\mid \Psi (0)>$ of the signal field to the state $\mid \Psi (t)>$ is
governed by
\begin{equation}
\mid \Psi (t)>\, = \,\exp [\kappa _{n}t(a^{n}-a^{\dagger n})]\mid \Psi
(0)>.
\end{equation}
The squeezing of this state was examined by Hillery, Zubairy and
W\'{o}dkiewicz \cite{11}. They showed that to any order in the coupling
constant $\kappa _{n},$ the vacuum state is not squeezed in the higher order
nonlinear optical processes $(n\geq 3).$

This important result stimulated the analysis of $n$th power squeezed
states. Interesting (and, generally, not yet completely explored) questions
arise in connection with this argument. Some of them are discussed in \cite
{5,13,19} and references therein.

Now let us make some comments about the Hamiltonian ${\cal H}_{4}.$ This
operator is closely related to ${\cal H}_{1}.$ This can be seen by means
of the phase transformation $a^{\prime }=ia,$ $a^{\prime \dagger
}=-ia^{\dagger },$ so that ${\cal H}_{4}$ takes the form
\begin{equation}
{\cal H}_{4}=-\frac{1}{\sqrt{2}}(a^{\prime 2\dagger }a^{\prime
}+a^{\prime \dagger }a^{\prime 2}+a^{\prime \dagger }+a^{\prime }).
\end{equation}
In other words, one has ${\cal H}_{4}=\frac{1}{2}{\cal H}_{1}$ (in
terms of the primed operators). This corresponds to pick up $q^{\prime }=-p$
and $p^{\prime }=q.$ In such a way ${\cal H}_{4}$ turns out to be the
inverse Fourier transform of (55). Therefore, the solutions of the
eigenvalue equation for the Hamiltonian operator ${\cal H}_{4}$ can be
derived from the solutions of the eigenvalue equation for the Hamiltonian
operator ${\cal H}_{1}$ (see (69)).

The eigenvalue equation for ${\cal H}_{1}$ ( ${\cal H}_{4}$ ) can be
investigated by means of the algebraic approach outlined in Section IX.

\section{ Equations and constants of motion related to the
operators ${\cal H}_{j}$}


The equations of motion for the Hamiltonians ${\cal H}_{j}$ $(j=1,2,3,4)$
arise immediately by using the Heisenberg representation. In other words, by
putting $a(t)=a(0)\exp (-it)$ and $a^{\dagger }(t)=a^{\dagger }(0)\exp (it)$
in the expressions (51) and (58), we easily find (as one expects) that $%
{\cal H}_{1}$ and ${\cal H}_{4}$ sastisfy the same equation of motion
(i.e., the equation for the harmonic oscillator of frequency $\lambda =1$):
\begin{equation}
\frac{d^{2}}{dt^{2}}{\cal H}_{1}+{\cal H}_{1}=0,\quad \frac{d^{2}}{%
dt^{2}}{\cal H}_{4}+{\cal H}_{4}=0.
\end{equation}
On the other hand, for the operators ${\cal H}_{3}=\frac{i}{\sqrt{2}}%
\frac{1}{3}(a^{3}-a^{\dagger 3})$ (see (57)) and ${\cal H}_{5}=\frac{1}{%
\sqrt{2}}\frac{1}{3}(a^{3}+a^{\dagger 3}),$ the same considerations made for 
${\cal H}_{1}$ and ${\cal H}_{4}$ in Section VII hold. One has that $%
{\cal H}_{3}$ and ${\cal H}_{5}$ obey the same equation of motion
(i.e., the equation for the harmonic oscillator of frequency $3$):
\begin{equation}
\frac{d^{2}}{dt^{2}}{\cal H}_{3}+9{\cal H}_{3}=0,\quad \frac{d^{2}}{%
dt^{2}}{\cal H}_{5}+9{\cal H}_{5}=0.
\end{equation}
Some comments on Eqs. (93) and (94) are presented in Section X.

At this point, we observe that in addition to the constants of motion
\begin{equation}
\stackrel{\sim }{q}=e^{-it\stackrel{\wedge }{H_{0}}}\stackrel{\wedge }{q}e^{it%
\stackrel{\wedge }{H_{0}}}=\stackrel{\wedge }{q}\cos t-\stackrel{\wedge }{p}%
\sin t,
\end{equation}
\begin{equation}
\stackrel{\sim }{p}=e^{-it\stackrel{\wedge }{H_{0}}}\stackrel{\wedge }{p}e^{it%
\stackrel{\wedge }{H_{0}}}=\stackrel{\wedge }{q}\sin t+\stackrel{\wedge }{p}%
\cos t,
\end{equation}
where $[\stackrel{\wedge }{q},\stackrel{\wedge }{p}]=i,$ $\stackrel{\wedge }{%
H_{0}}=\frac{1}{2}(\stackrel{\wedge }{p}^{2}+\stackrel{\wedge }{q}^{2})$ and $%
\stackrel{\sim }{q}=-\frac{1}{2}I_{2}^{(2)},$ $\stackrel{\sim }{p}=\frac{1}{2}%
I_{2}^{(1)},$ we obtain also the following set of constants:
\begin{equation}
\stackrel{\sim }{H_{1}}={\cal H}_{1}\cos t-2{\cal H}_{4}\sin t,
\end{equation}
\begin{equation}
\stackrel{\sim }{H_{4}}=\frac{1}{2}{\cal H}_{1}\sin t+{\cal H}_{4}\cos
t,
\end{equation}
\begin{equation}
\stackrel{\sim }{H}_{3}={\cal H}_{3}\cos 3t-{\cal H}_{5}\sin 3t,
\end{equation}
\begin{equation}
\stackrel{\sim }{H}_{5}={\cal H}_{3}\sin 3t+{\cal H}_{5}\cos 3t.
\end{equation}
In terms of the arbitrary constants $\stackrel{\sim }{H}_{j}(j=1,4,3,5),$ one
can express the general solutions of Eqs. (93) and (94), which can be easily
found by inverting the transformations (97), (98) and (99), (100),
respectively.


\section{A possible algebraic framework for the study of Eq. (69)}


For $\varepsilon \neq 0,$ Eq. (69) could be investigated following different
analytical techniques. Among these, an important role is played by the
algebraic approach, which allows one to express Eq. (69) in terms of the
generators of a \textit{quadratic} algebra \cite{20,21}.

As an example of a quadratic algebra, we can consider a nonlinear
deformation of the (linear) $su(1,1)$ algebra, defined by the relations

\begin{equation}
\lbrack J_{0},J_{\pm }]=\pm J_{\pm },
\end{equation}
\begin{equation}
\lbrack J_{+},J_{-}]=P(J_{0}),
\end{equation}
where the Jacobi identity holds. $J_{\pm }$ are the ladder operators, and $%
P(J_{0})$ is a second-degree polynomial function of the diagonal operator $%
J_{0}.$ $P(J_{0})$ can be written in the form
\begin{equation}
P(J_{0})=\alpha _{1}+\alpha _{2}J_{0}+\alpha _{3}J_{0}^{2},
\end{equation}
where $\alpha _{i}$ ($i=1,2,3)$ are arbitrary coefficients ($\alpha _{3}\neq
0).$

Now let us introduce the (bosonic) realization
\begin{equation}
J_{0}=a^{\dagger }a+\frac{1}{2},
\end{equation}
\begin{equation}
J_{-}=-(a^{\dagger }a^{2}+a),
\end{equation}
\begin{equation}
J_{+}=a^{\dagger 2}a+a^{\dagger },
\end{equation}
with $J_{+}=-(J_{-})^{\dagger }.$ Then, the operators (104)--(106) turn out
to obey the commutation relations
\begin{equation}
\lbrack J_{0},J_{\pm }]=\pm J_{\pm },\quad [J_{+},J_{-}]=\frac{1}{4}%
+3J_{0}^{2}.
\end{equation}
We remark that the quadratic algebra (107) is a finite $W_{3}^{(2)}-$algebra 
\cite{20,21}, which corresponds to choose $\alpha _{1}=\frac{1}{4},$ $\alpha
_{2}=0,$ and $\alpha _{3}=3$ in (103).

By virtue of (104)--(106), Eq. (69) can be re-expressed as
\begin{equation}
\Omega (\varepsilon )\mid \chi >\quad =0,
\end{equation}
with
\begin{equation}
\Omega (\varepsilon )=J_{+}-J_{-}-\varepsilon _{.}
\end{equation}
Therefore, the eigenvalue problem (69) can be formulated in terms of the
generators $J_{\pm }$ of a quadratic algebra of the type $W_{3}^{(2)}.$ The
Casimir operator of the quadratic algebra (107) is considered in the
Appendix.

\section{ Conclusions}

In the context of the existence of an infinite set of fouled Lagrangians and
Hamiltonians for the generalized (time-dependent) oscillator, we have
studied a special case where the frequency of the oscillator is assumed to
be a constant $\lambda $ (harmonic oscillator). By way of example, we have
considered a pair of independent fouled Lagrangians $(L_{2}^{(1)}$, $%
L_{2}^{(2)})$ (see (23)-(24)) and, in correspondence, a pair of fouled
independent Hamiltonians $(K_{1},$ $K_{2})$ (see (39)-(40)) which lead, at
the classical level, to the same equation of motion provided by the
conventional Lagrangian and Hamiltonian. Both $(L_{2}^{(1)}$, $L_{2}^{(2)})$
and $(K_{1},$ $K_{2})$ are explicitly time-dependent.The method followed to
find these alternative Lagrangians and Hamiltonians implies the construction
of two independent invariants (constants of motion), $I_{2}^{(1)}$ and $%
I_{2}^{(2)};$ in terms of them the canonical variables $q$ and $p$ can be
expressed. These invariants, which are of the N\"{o}ther type \cite{22}, are
connected with the ''quadrature-phase amplitudes'' appearing in the problem
of generation of squeezed states in certain optical devices \cite{23}.

In this paper we have focused our attention mainly on the quantized version
of the fouled Hamiltonians $(K_{1},$ $K_{2}).$ Since $K_{1}$ and $K_{2}$ are
related to the cubic polynomials $H_{1}=\sqrt{\lambda }(p^{2}$ $+\lambda
^{2}q)q$ and $H_{2}=\frac{2}{3\sqrt{\lambda }}p^{3}$ by a formal rotation,
the canonical quantization prescription affects essentially $H_{1}$ and $%
H_{2}.$ Our purpose has been to study, at the quantum level, the
Hamiltonians ${\cal H}_{1}$ and ${\cal H}_{2}$ (see (51), (52))
corresponding to $H_{1}$ and $H_{2}$ . The model described by ${\cal H}%
_{2}$ is associated with an eigenvalue problem given by a (linear) third
order differential equation, with constant coefficients, which is exactly
solvable \cite{9}. Furthermore, we have put ${\cal H}_{2}={\cal H}_{3}+%
{\cal H}_{4}$ , where ${\cal H}_{4}$ (see (58)) turns out to be
closely related to the quantum model ${\cal H}_{1}$ in the sense
discussed in Section VII. On the other hand, the quantum model ${\cal H}%
_{3}$ has a well-defined physical interpretation. It belongs to a class of
Hamiltonians which finds applications in the field of $n$th power squeezed
states.

Finally, the operator ${\cal H}_{1}$ , considered in Section VI, has
deficiency indices (1,1) and allows a one-parameter family of self-adjoint
extensions, each having a purely discrete spectrum on the real line. The
spectra of two different extensions have no point in common. Since different
self-adjoint extensions correspond to different dynamics, we needed to fix a
given dynamics. This has been carried out by choosing the value $\varepsilon
=0$ for the eigenvalue parameter. In this case all the differential
equations coming from (69) in all the representations: $n-$ rep, $q-$ rep
and $z-$ rep, can be solved exactly. In this case, square-integrable
solutions of the eigenvalue equations are explicitly determined.

In order to investigate some properties of the Hamiltonians ${\cal H}_{j}$
($j=1,2,3,4),$ in Section VIII we have introduced the operator ${\cal H}%
_{5}\sim (a^{3}+a^{3\dagger }),$ which is connected with ${\cal H}_{3}$,
as one can see by using the transformation $a^{\prime }=ia,$ $a^{\prime
\dagger }=-ia^{\dagger }.$ We have built up a set of constants of motion
involving $({\cal H}_{1},{\cal H}_{4})$ and $({\cal H}_{3},{\cal %
H}_{5}).$ These constants play the role of the arbitrary constants present
in the general solution of two equations of the harmonic oscillator-type of
frequencies $1$ and $3,$ respectively$,$ which are satisfied by $({\cal H}%
_{1},{\cal H}_{4})$ and $({\cal H}_{3},{\cal H}_{5})$ (see (93) and
(94)).

To conclude, we point out that quadratic (and, more in general, polynomially
deformed algebras) take place in quantum optics in relation to the
construction of coherent states and in the description of multiphoton
processes (see, for examples, \cite{24,25} and references therein). Keeping
in mind these problems, it should be of interest to deal with the
quantization of fouled Hamiltonians (of the generalized oscillator)
expressed by polynomials in the canonical variables $(q,p)$ of degree higher
than three.


{\LARGE Acknowledgments}

We would like to express our gratitude to the (anonymous) referee for
notable suggestions, especially in regard to the improvement of Section VI.
We also thank Dr. G. Esposito, of Istituto Nazionale di Fisica Nucleare,
Sezione di Napoli, for helpful discussions.

\section{ Appendix: the Casimir invariant for the quadratic algebra (107)}


We remind the reader that a standard form of a quadratic algebra, which is
important in the treatment of coherent states of trilinear boson
Hamiltonians \cite{26}, is \cite{24,25,27}
\begin{equation}
\lbrack N_{0},N_{\pm }]=\pm N_{\pm },\quad [N_{+},N_{-}]=\pm 2N_{0}+\delta
N_{0}^{2},  \lab{A1}
\end{equation}
where the positive (negative) sign of $2N_{0}$ indicates a polynomially
deformed $su(2)$ $(su(1,1)),$ and $\delta $ is a parameter.

The Casimir operator is given by \cite{24,25,27}
\begin{equation}
C=N_{-}N_{+}+N_{0}(N_{0}+1)[1+\frac{\delta }{6}(2N_{0}+1)].  \lab{A2}
\end{equation}
By setting
\begin{equation}
J_{0}=a_{0}N_{0}+b_{0},  \lab{A3}
\end{equation}
\begin{equation}
J_{-}=k_{1}N_{-},  \lab{A4}
\end{equation}
\begin{equation}
J_{+}=k_{2}N_{+},  \lab{A5}
\end{equation}
and choosing, for example,
\begin{equation}
a_{0}=1,\quad b_{0}=\frac{i}{2\sqrt{3}},\quad k_{1}k_{2}=\frac{i\sqrt{3}%
}{2},  \lab{A6}
\end{equation}
the commutator relations (107) are converted into
\begin{equation}
\lbrack N_{0},N_{\pm }]=\pm N_{\pm },\quad [N_{+},N_{-}]=2N_{0}+\delta
N_{0}^{2},  \lab{A7}
\end{equation}
with $\delta =-2i\sqrt{3}.$ Hence, the Casimir operator (\ref{A2}) reads
\begin{equation}
\quad \quad C=N_{-}N_{+}+N_{0}(N_{0}+1)[1-\frac{i}{\sqrt{3}}%
(2N_{0}+1)]. \lab{A8}
\end{equation}

\end{document}